# Spin State *versus* Potential of Zero Charge as Predictors of Density-Dependent Oxygen Reduction in M–N–C Electrocatalysts

Di Zhang[1,2], Zixun Yu[3], Fangzhou Liu[3], Yumeng Li[3], Jiaxiang Chen[3], Xun Geng[4], Yuan Chen[3], Li Wei[3,*], and Hao Li[1,*]

[1] Advanced Institute for Materials Research (WPI-AIMR), Tohoku University, Sendai, 980-8577, Japan

[2] The Frontier Research Institute for Interdisciplinary Sciences (FRIS), Tohoku University, Sendai, 980-8577, Japan

[3] School of Chemical and Biomolecular Engineering, The University of Sydney, Darlington, New South Wales, 2008, Australia

[4] Sydney Analytical, Core Research Facilities, The University of Sydney, New South Wales, 2008, Australia

* Corresponding Authors:

Email: l.wei@sydney.edu.au (L. W.)

Email: li.hao.b8@tohoku.ac.jp (H. L.)

**Abstract**

Metal-site density strongly influences oxygen reduction activity and selectivity in M–N–C electrocatalysts, but the descriptors that predict these trends remain under debate. Here, we compare spin state and the potential of zero charge (PZC) as predictors of density-dependent oxygen reduction behavior in Fe–N–C and Co–N–C catalysts. Using constrained-magnetization calculations combined with Landau analysis, we find that the ground-state magnetic moments vary only weakly across a broad range of metal-site densities, suggesting that magnetic descriptors alone cannot account for the pronounced performance changes. In contrast, explicit-solvent simulations reveal systematic density-dependent shifts in PZC, which alter the interfacial electric field and thereby modulate field-sensitive adsorption energetics of ORR intermediates. Incorporating these PZC shifts into a pH–field-coupled microkinetic model captures the density-dependent activity trends and reproduces the experimentally observed increase in two-electron selectivity at lower site densities under acidic conditions. Experimental PZC measurements further support the predicted trend. Within the investigated Co–N–C and Fe–N–C series, PZC-dependent electrostatics consistently rationalize the density-dependent ORR trends, whereas the examined magnetic descriptors show comparatively weak variation.

## Introduction

Metal–nitrogen–carbon (M–N–C) electrocatalysts have emerged as an important class of oxygen reduction reaction (ORR) catalysts because their atomically dispersed metal centers offer opportunities to tune activity and reaction selectivity through local coordination and electronic structure engineering.[1-3] In addition to the identity of the metal center itself, the density of metal sites has increasingly been recognized as a key structural variable in these materials.[4, 5] However, the physicochemical descriptor that best predicts these density-dependent catalytic trends remains unresolved.

One possible explanation is that changing the average distance between neighboring metal sites modifies the local electronic structure of the active center, including its spin state or magnetic moment.[4-10] Spin-related effects have attracted broad attention in electrocatalysis, particularly in systems where spin populations are actively perturbed by external magnetic fields[11-13] or by deliberately engineered coordination environments.[8, 14-23] For example, magnetic moments have been used to guide the design of Co single-atom catalysts through second-coordination-shell engineering,[24] while increased Fe loading has been associated with spin-state transitions and altered $H_2O/O_2$ exchange kinetics in Fe–N–C catalysts.[25] These findings highlight the importance of spin in electrocatalysis and motivate a closer examination of its role in density-tuned M–N–C materials. Indeed, spin descriptors have also been invoked to rationalize activity variations in field-free M–N–C catalysts with different site densities. [4, 5] However, whether changes in site density systematically alter the lowest-energy spin state remains an open question, particularly because density functional theory calculations can converge to metastable magnetic solutions depending on the initial magnetic configuration and convergence pathway.

At the same time, changing metal-site density can influence the electrochemical interface through variations in local charge distribution, solvent organization, and electrostatic interactions. The broader potential of electrostatic descriptors has been illustrated by electrostatic-potential-guided design of atomic-pair catalysts.[26] For aqueous M–N–C interfaces, the PZC is particularly relevant because it governs the interfacial electric field at a fixed electrode potential and thereby affects the stabilization of field-sensitive adsorbates.[27] For ORR, where the energetics of key intermediates depend on pH, [28] electrode potential, and interfacial field effects, a density-dependent PZC shift may influence both catalytic activity and reaction-pathway selectivity. [1, 2] These considerations motivate a direct comparison of magnetic and interfacial electrostatic descriptors across the metal-site-density series.

Here, we compare spin state and the potential of zero charge as predictors of density-dependent oxygen reduction in Fe–N–C and Co–N–C electrocatalysts under field-free conditions. We first develop a rapid constrained-magnetization protocol to evaluate the magnetic ground states of M–N–C catalysts across a range of metal-site densities. By mapping the energy as a function of magnetization and analyzing the resulting energy landscape with a Landau expansion,[29] we assess whether ground-state magnetic moments vary systematically with site density. We then evaluate the PZC using explicit-solvent models of the electrochemical interface and examine how density-dependent PZC shifts influence ORR energetics through a pH–field-coupled microkinetic framework.[30] Our calculations show that the ground-state magnetic moments of Fe–N–C and Co–N–C catalysts change only weakly with site density, whereas the PZC shifts systematically across the density series. Incorporating these PZC shifts into the microkinetic model captures the density-dependent trends in ORR activity and reproduces the experimentally observed increase in two-electron selectivity at lower site densities under acidic conditions. Experimental measurements of spin-sensitive X-ray emission spectra, PZC, and ORR performance further support this comparison. Together, these findings establish PZC as a more effective predictor than spin state for density-dependent oxygen reduction in the M–N–C systems investigated here, highlighting the critical role of interfacial electrostatics in governing catalytic activity and selectivity.

## Methods

### Computational Methods

Spin-polarized DFT calculations were performed using VASP with the RPBE[31, 32] exchange–correlation functional and PAW potentials (plane-wave cutoff 520 eV), including D3[33] dispersion and dipole corrections. Adsorption free energies were evaluated using the computational hydrogen electrode (CHE) with zero-point energy, entropic, and solvation corrections. External electric-field effects (−0.6 to 1.0 V $Å^{-1}$) were computed using VASP and parameterized to obtain dipole moments and polarizabilities for potential- and pH-dependent free-energy corrections. PZCs were determined from explicit-solvent ab initio molecular dynamics (AIMD) simulations and extracted from work-function analysis referenced to the standard hydrogen electrode (SHE)[27]. ORR activity and selectivity were assessed using CatMAP-based[34] microkinetic modeling combining catalyst-specific adsorption free energies, electric-field responses, and calculated PZCs with literature-parameterized elementary-step kinetics. Both $4e^-$ and $2e^-$ pathways were included. The model is used to compare density-dependent trends rather than to provide a fully first-principles kinetic description of every elementary step. The kinetic parameters and their sources are summarized in Supplementary Table S5. A representative explicit-solvent HOO* cleavage calculation was additionally performed using constrained AIMD and Blue Moon thermodynamic integration (Supplementary Section 1.7 and Fig. S1).[35] Additional computational details are provided in the **Supplementary Information**, including DFT settings and adsorption free-energy calculations (**Section 1.1, Table S1**), constrained-magnetization calculations (**Section 1.2**), electric-field calculations and pH/potential-dependent free-energy corrections (**Section 1.3, Fig. S10**), the explicit-solvent AIMD protocol and PZC evaluation (**Section 1.4**), and CatMAP-based ORR microkinetic modeling (**Section 1.5, Reactions 13–20**). The convergence tests of force threshold and plane-wave cutoff are documented in Supplementary Section 1.8 and Figs. S2–S3.

## Experimental Section

**Materials and catalyst synthesis.** Co– and Fe–N–C catalysts with controlled metal-site densities were prepared using a hydrogel-anchoring strategy with minor modifications (details in Supplementary Information, Sections 2.1–2.2). Briefly, polypyrrole (PPy) hydrogels were impregnated with methanolic solutions of metal acetylacetonate precursors at three concentrations (0.1, 0.01, and 0.001 M), followed by drying and two-step pyrolysis at 900 °C under Ar with an intermediate acid leaching step to remove unstable metal species. The resulting catalysts are denoted as Hi, Mid, and Low M–N–C (M = Co or Fe) according to the precursor concentration used (**Supplementary Information, Section 2.2**).

Structural characterization. Aberration-corrected HAADF-STEM was used to visualize isolated metal sites in multiple independently selected regions of each sample; atomic-scale bright features were identified by an automated Laplacian-of-Gaussian procedure with identical parameters for all images, and their areal density and projected nearest-neighbor distances were evaluated in the thin regions of the support (9 regions per metal, Supplementary Information Section 2.3, Table S2). Metal loadings were determined by ICP-OES after aqua regia digestion. Co/Fe K-edge XAS was collected in fluorescence mode at the Australian Synchrotron (MEX-1), with the Mid-density samples re-measured on three independent spots; all spectra were calibrated against the simultaneously measured metal foil, and XANES, EXAFS (M–$N_4$ model) and wavelet-transform analyses were performed as described in Section 2.3. Kβ XES and $^{57}$Fe Mössbauer spectroscopy were used to probe the local spin state (Sections 2.6–2.7). Wettability, pore structure and $H_2O_2$ reduction were characterised as described in Sections 2.8–2.10.

Electrochemical measurements. ORR performance was evaluated in a three-electrode rotating ring-disk electrode (RRDE) configuration in $O_2$-saturated 0.1 M $HClO_4$ at 25 °C. Catalyst inks were drop-cast onto a glassy carbon disk (0.1 mg cm$^{-2}$). LSV was recorded at 10 mV s$^{-1}$ and 1600 rpm, while the Pt ring was held at 1.25 V vs RHE to quantify $H_2O_2$. The $H_2O_2$ selectivity $XH_2O_2$ = 200·(iring/N)/(idisk + iring/N) and the electron-transfer number n = 4·idisk/(idisk + iring/N) were calculated with the experimentally calibrated ring collection efficiency N = 24.5 ± 0.3 % (Supplementary Information, Section 2.4); kinetic current, mass activity and TOF were obtained from Koutecký–Levich analysis as described there.

Experimental PZC determination. The PZCs were determined by electrochemical impedance spectroscopy (100 kHz–0.1 Hz, 5 mV rms) in Ar-saturated 0.1 M $HClO_4$ at 15 DC potentials, by extracting the double-layer capacitance (Cdl = −1/(2πf·Z″) at 1 Hz) as a function of potential; the PZC was assigned to the minimum of the E–Cdl curve refined by a parabolic fit, and each measurement was repeated three times (Supplementary Information, Section 2.5, Table S9).

## Results and Discussion

### Spin State Shows Weak Dependence on Metal-Site Density

Previous studies have suggested that in M–N–C catalysts, reducing the distance between metal atoms [4] or enhancing the metal site densities [5] can modify their spin states through inter-site magnetic interactions, thereby giving rise to density-dependent catalytic activity. To examine whether magnetic moments vary with the spatial distribution of metal sites, we first performed constrained-magnetization calculations for M–N–C models with varying metal–metal separations at a fixed metal-site number. These calculations probe the effect of inter-site spacing. We additionally varied the number of Co sites within the same in-plane cell area to test the effect of areal site density directly, as discussed below. As illustrated in **Fig. 1a**, we applied controlled magnetic-moment perturbations to identical structural models by varying both the target magnetization (μ = 0-3 μB) and the penalty strength λ (0, 10, 100) [36], allowing the total energy to be mapped as a function of magnetization. **Fig. 1b** summarizes the results of constrained-magnetization calculations for $Co_2$–N–C and $Fe_2$–N–C, where different initial magnetic moments and penalty strengths were systematically examined using the same structural model. In the absence of magnetic constraints (penalty strength = 0), both systems consistently converge to ground magnetic states, with local magnetic moments of approximately 0.87 μB per Co atom and 1.94 μB per Fe atom, regardless of the initial magnetic configuration. Increasing the penalty strength allows access to alternative magnetic states with substantially different magnetic moments. For instance, at a penalty strength of 100 and a target moment of 3 μB per metal atom, the final magnetic moments reach 4.00 μB per Co atom and 4.14 μB per Fe atom. By sampling multiple magnetic initializations and constraint strengths, this protocol reduces reliance on any single self-consistent magnetic solution. Details of the constraint implementation and the treatment of zero-target moments are provided in **Supplementary Section 1.2**.

To evaluate whether inter-site separation influences the preferred magnetization, we examined $Co_2$–N–C and $Fe_2$–N–C models with metal–metal distances of 8.6, 12.9, 13.8, and 17.2 Å (**Fig. 1c**). The relative energies calculated at different total cell magnetizations are presented in Fig. 1d,e. These energy–magnetization profiles are well described by the Landau expansion, [29] $E(M) = aM^2 + bM^4 + E_0$ (see **Fig. S5** for a schematic illustration). The fits yield $R^2 = 0.999$ for all four $Co_2$–N–C structures and $R^2 = 0.989$–0.995 for $Fe_2$–N–C, allowing the magnetizations at the fitted energy minima to be extracted. For $Co_2$–N–C, M_min varies only from 1.958 to 1.995 μB per cell, while $Fe_2$–N–C exhibits an equally narrow range of 4.219–4.244 μB per cell. These fitted ranges correspond to variations of only 0.037 μB per cell for $Co_2$–N–C and 0.025 μB per cell for $Fe_2$–N–C. The weak variation of these energy-minimizing magnetizations across the investigated distances indicates that changes in proximity between metal centers do not substantially alter their preferred magnetic states under field-free conditions.

To assess whether adsorption modifies this spacing dependence, we examined HO*-adsorbed $Co_2$–N–C and $Fe_2$–N–C structures across the same set of inter-site separations (**Fig. S6**). Although HO* adsorption shifts the magnetization associated with the energy minimum relative to the pristine surfaces, neither adsorbed series exhibits a pronounced trend with metal–metal separation.

We next tested whether this weak dependence persists when the site number or local coordination motif is changed. Increasing the number of isolated Co–$N_4$ sites from 3 to 6 within the same in-plane cell area produces similar mean local Co moments at the lowest sampled corrected energies, despite the twofold variation in areal site density (**Fig. S7**). Pyrrolic-type Fe–$N_4$ models likewise yield mean local Fe moments close to 1.38 μB per Fe across the high-, medium-, and low-density series, consistent with the weak density dependence found for the pyridinic-type models (**Fig. S8**). Across these controls, the preferred magnetic moments show only weak dependence on site spacing or density within the models examined.

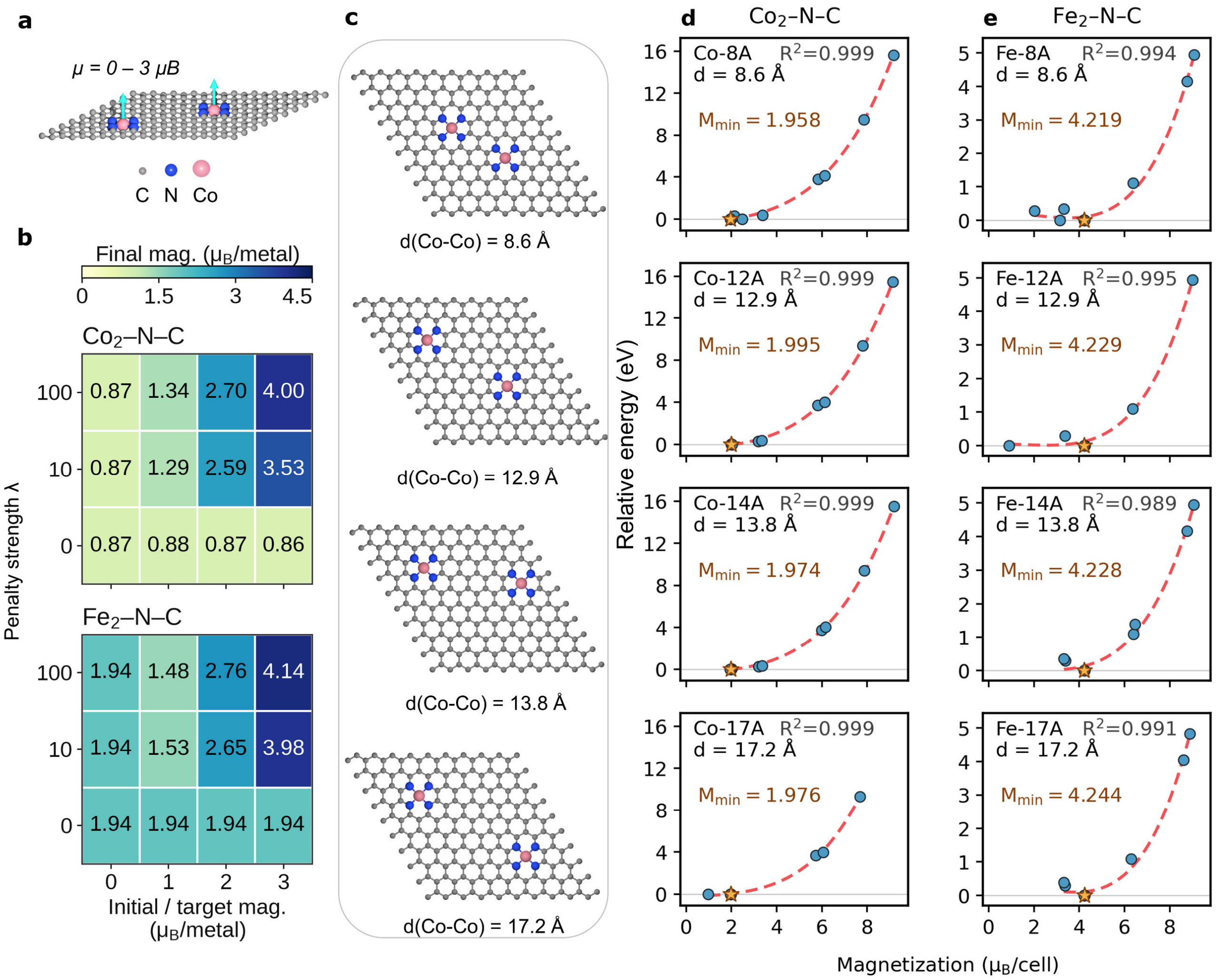

**Fig. 1. Systematic magnetic-moment screening and energy–magnetization relationships of M–N–C models at different metal–metal separations. (a)** Schematic illustration of the constrained-magnetization protocol with initial/target magnetic moments ranging from 0 to 3 μB per metal site. **(b)** Final converged magnetic moments (μB per metal) for $Co_2$–N–C and $Fe_2$–N–C at different initial/target moments and penalty strengths (λ = 0, 10, and 100). (c) Representative dual-site models with metal–metal separations of 8.6, 12.9, 13.8, and 17.2 Å ($Co_2$–N–C structures shown). The corresponding $Fe_2$–N–C models share identical atomic configurations, with Co atoms replaced by Fe atoms. **(d, e)** Calculated relative energies as functions of the total cell magnetization for $Co_2$–N–C **(d)** and $Fe_2$–N–C **(e)**, respectively. Blue circles are the calculated results, dashed red curves are Landau fits, and orange stars identify the fitted energy minima. Corresponding results for HO*-adsorbed models are provided in **Fig. S6.**

**Potential of Zero Charge Shifts Systematically with Metal-Site Density**

If the magnetic ground state is largely insensitive to metal-site density and inter-site separation, the origin of the pronounced performance variations observed in M–N–C catalysts remains an open question. To address this issue, we constructed a series of Co–N–C and Fe–N–C models with systematically varied metal-site densities, denoted as high-, medium-, and low-density configurations (**Hi**, **Mid**, and **Low**; **Fig. 2a**). For each model, the PZC was evaluated using an explicit solvent representation of the electrochemical interface (**Fig. 2b**). *Ab initio* molecular dynamics (AIMD) simulations were performed, and 100 snapshots were extracted to sample interfacial fluctuations [27, 37]. PZC values were then computed for each snapshot following the procedure described in the **Supplementary Information 1.4**. The resulting PZC distributions exhibit Gaussian-like profiles (**Fig. 2b**), indicating well-defined ensemble-averaged characteristics for each catalyst.

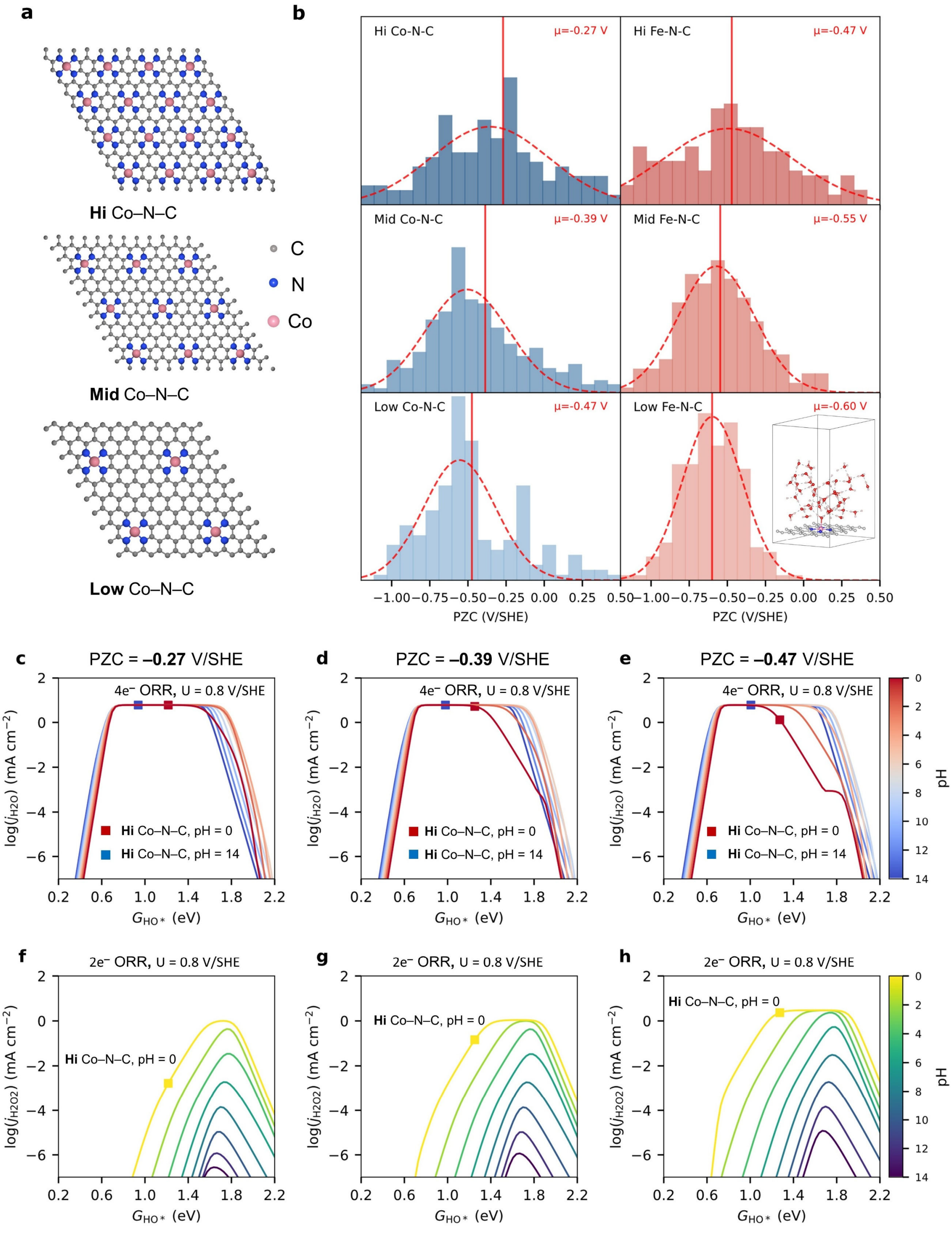

**Fig. 2. Density-dependent electrochemical origins of activity and selectivity in M–N–C catalysts.** (**a**) Structural models of Co–N–C catalysts with high, medium, and low metal-site densities. (**b**) Explicit-solvent interfacial model used for PZC calculations, with ensemble averaging over AIMD trajectories. (**c-**

**e**) pH-dependent activity volcano plots for the 4-electron ORR pathway of Co–N–C catalysts under different PZC conditions obtained from the pH–electric-field-coupled microkinetic model. (**f–h**) Corresponding volcano plots for the 2-electron ORR pathway under the same conditions. As the metal-site density decreases, the PZC shifts to more negative values, (**c, f**) PZC = −0.27 V vs SHE, (**d, g**) PZC = −0.39 V vs SHE, (**e, h**) PZC = −0.47 V vs SHE, leading to a systematic suppression of 4-electron ORR activity and a concurrent enhancement of the 2-electron pathway in an acidic solution. All volcano plots in panels (c–h) are evaluated at U = 0.8 V vs SHE over pH 0–14; the indicated PZCs are referenced to SHE. These conditions are distinct from the controlled PZC-only sensitivity analysis at pH = 0 and U = 0.30 V vs RHE (Supplementary Section 1.6 and Fig. S17).

Analysis of the mean PZC values reveals a clear and systematic dependence on metal-site density. For Co–N–C catalysts, the average PZC shifts from −0.27 V to −0.39 V and −0.47 V versus standard hydrogen electrode (SHE) as the site density decreases from high to medium and low, respectively. A similar trend is observed for Fe–N–C systems, with mean PZC values of −0.47 V, −0.55 V, and −0.60 V versus SHE for the high-, medium-, and low-density models. These monotonic shifts indicate that reducing metal-site density leads to a progressive lowering of the PZC, pointing to a density-driven modification of the electrochemical boundary conditions at the catalyst–electrolyte interface.

To distinguish PZC effects from other density-dependent energetic contributions, we first compared the zero-field adsorption energetics and intrinsic field responses of the key ORR intermediates across the Co–N–C density series. The adsorption free energies calculated without interfacial electric-field corrections are similar among the Hi-, Mid-, and Low-density models (**Fig. S9**), as are their intrinsic field-response coefficients (**Fig. S10**). These comparisons suggest that substantial changes in zero-field binding or intrinsic field sensitivity are not the main source of the predicted density dependence. We therefore examined how the PZC shifts alter the actual interfacial field and, consequently, intermediate adsorption energetics at a common applied potential.

Within the Helmholtz-layer approximation, the interfacial field depends on the electrode potential relative to its PZC, $E = C_H(U_{SHE} - U_{PZC})/(\varepsilon\varepsilon_0)$. At fixed applied potential and interfacial capacitance and permittivity, the negative PZC shift across the present density series therefore strengthens the local electric field (**Supplementary Sections 1.3 and 1.6**). At a fixed RHE-referenced potential, the SHE-referenced potential also varies with pH according to U_SHE = U_RHE − 0.0591 pH at 298 K, allowing the same electrostatic framework to be applied across different electrolyte conditions. Such changes in the local electric field can affect the adsorption energetics of ORR intermediates differently, potentially altering the competition between the $2e^-$ and $4e^-$ pathways. To isolate the contribution of PZC, we performed a controlled analysis at pH = 0 and U = 0.30 V vs RHE, varying only the calculated PZC while keeping the zero-field adsorption energies and field-response coefficients fixed to those of the Mid-density Co–N–C model. Shifting the PZC from the Hi- to Low-density value stabilizes $H_2O_2$* by 0.212 eV and destabilizes O* by 0.085 eV. The resulting 0.297 eV decrease in $\Delta G(H_2O_2^*) - \Delta G(O^*)$ indicates a thermodynamic shift toward the $2e^-$ pathway (**Supplementary Section 1.6 and Fig. S17**).

To further examine how these PZC-dependent energetic changes influence ORR activity and selectivity, we employed the pH–electric-field-coupled microkinetic model. [1, 2, 28, 30] The resulting pH-dependent activity volcano plots for the 4-electron and 2-electron ORR pathways of Co–N–C catalysts are presented in Fig. 2c–e and Fig. 2f–h, respectively (corresponding results for Fe–N–C are shown in **Fig. S11**). Under acidic conditions (pH = 0), the predicted site-specific 4-electron ORR activity decreases systematically as the PZC shifts to more negative values (red square markers in Fig. 2c–e). In contrast, the 2-electron ORR activity exhibits the opposite trend, increasing monotonically with decreasing PZC (yellow square markers in Fig. 2f–h). These results indicate that lowering the metal-site density in M–N–C catalysts under the acidic conditions investigated here leads to a redistribution of ORR pathways: the intrinsic 4-electron activity of individual sites is progressively suppressed, while the 2-electron pathway is promoted. Consequently, the total current density decreases with decreasing site density, whereas the selectivity toward the 2-electron ORR increases. The predicted trend in overall activity is consistent with previous experimental observations.[4] The accompanying increase in $H_2O_2$ selectivity at lower site densities is examined experimentally in the following section.

**Experimental validation of density-dependent ORR selectivity**

To validate our theoretical predictions, particularly the increased 2e-ORR selectivity of lower site density under acidic conditions, we conducted systematic catalyst synthesis, characterizations, and electrochemical assessment. Co–N–C and Fe–N–C catalysts with different metal loadings were prepared following a hydrogel-based method reported earlier (see details in the Experimental Section)[4]. Representative high-angle annular dark-field scanning transmission electron microscope images (HAADF-STEM) of Co–N–C catalysts with high (Hi Co-N-C), medium (Mid Co-N-C), and low (Low Co-N-C) metal site densities are displayed in **Fig. 3a-3c** (corresponding images of Fe–N–C catalysts are shown in **Figure S12a-c**). Discrete atomic-scale bright features are observed throughout the examined regions, without resolvable nanoparticles or extended metal-rich domains (**Fig. 3a–c**; marked images and magnified views in **Fig. S21**). To quantify the density variation objectively, the bright features were identified using an automated detection procedure applied with identical parameters to all images (**Supplementary Information 2.3**). The areal density of detected features in the thin regions of the support decreases from 2.03 to 0.65 and 0.38 $nm^{-2}$ for Hi, Mid and Low Co–N–C (**Fig. 3d**), consistent with the systematic decrease in Co loading from 6.7 to 2.9 and 0.6 wt.% determined by inductively coupled plasma optical emission spectroscopy (ICP-OES); the corresponding mean projected nearest-neighbor distances increase from 0.40 to 0.56 and 0.76 nm. The individual-region values shown in the inset of Fig. 3d illustrate the spatial variability in the detected feature density within each sample. The Fe–N–C series shows the same trend (2.24, 1.06 and 0.13 $nm^{-2}$; 0.38, 0.43 and 0.89 nm; Fig. S12d and Table S2). Together, the multi-region imaging and statistical analysis support the designed variation in the density of predominantly atomically dispersed metal sites.

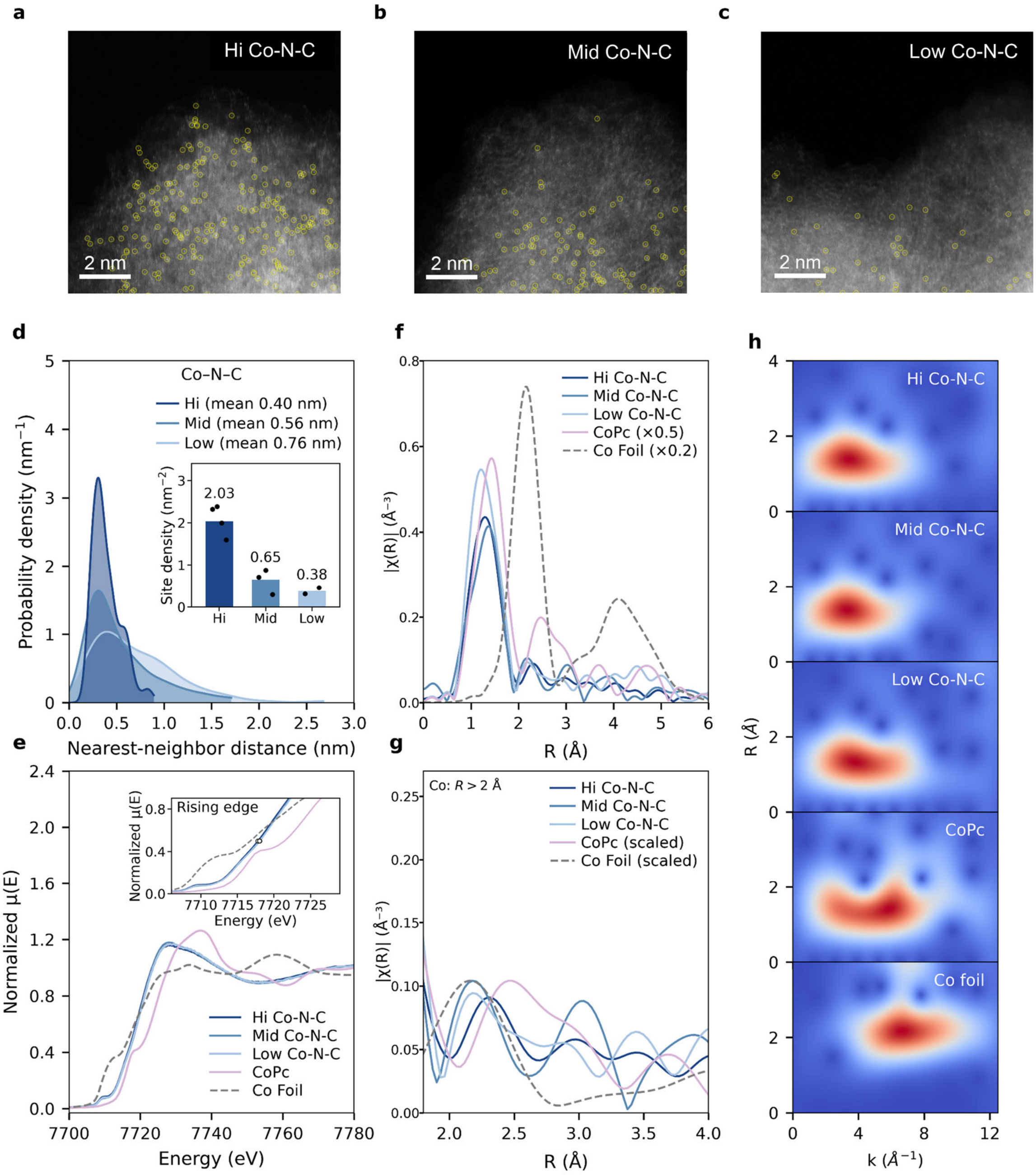

**Fig. 3. Structural characterization of Co–N–C catalysts with different metal-site densities.** (a–c) HAADF-STEM images of Hi-, Mid-, and Low-density samples. Yellow circles mark automatically detected bright features; scale bars: 2 nm. (d) Projected nearest-neighbor-distance distributions; the inset shows feature densities and individual-region values. (e) Co K-edge XANES spectra with CoPc and Co-foil references; inset: rising-edge region. (f) Fourier-transformed $k^2$-weighted EXAFS spectra, with CoPc and Co-foil amplitudes multiplied by 0.5 and 0.2. (g) Enlarged higher-R comparison with rescaled references. (h) WT-EXAFS maps of the samples and references. Reference scaling facilitates visual comparison and is not used to quantify metallic fractions.

To further examine the local coordination environment, Co and Fe K-edge XAS measurements were performed (Fig. 3e). The inset of Fig. 3e highlights the rising-edge region for closer comparison among the catalysts and references. The overall Co K-edge XANES profiles resemble that of CoPc and differ markedly from that of Co foil, supporting a predominantly Co–N-coordinated local structure. The Fourier-transformed EXAFS spectra exhibit a dominant first-shell peak near 1.5 Å, assigned to Co–N scattering (Fig. 3f). Weak higher-R features are also present, but their radial positions alone do not establish Co–Co coordination. Similar features in the molecular CoPc reference can arise from higher-shell and multiple-scattering contributions (Fig. 3g). The WT-EXAFS maps likewise show predominantly low-k first-shell contributions without a distinct foil-like Co–Co maximum (Fig. 3h). Additional fits incorporating a metallic M–M path yield near-zero coordination numbers under the stated fitting constraints (Table S6). The corresponding Fe results, k-space comparisons and fitting results are provided in Figs. S12–S14 and Tables S3 and S6. Together, these measurements support predominantly atomically dispersed $M–N_4$ sites, while trace clustered species below the sensitivity of the measurements cannot be excluded.

To experimentally verify that the local magnetic character of Fe/Co–N–C catalysts remains nearly invariant with site density, $K_\beta$ X-ray emission spectroscopy (XES) was employed to probe the local electronic structure and spin state of the metal centers in Hi/Mid/Low Co–N–C and Fe–N–C. As shown in **Fig. S15a–f**, both Co and Fe Kβ emission spectra exhibit highly similar overall line shapes across the entire density series. Peak deconvolution resolves the main $K\beta_{1,3}$ feature and the low-energy $K_{\beta'}$ satellite, where the $K\beta'/K\beta_{1,3}$ area ratio serves as a spin-sensitive descriptor. The $K\beta'/K\beta_{1,3}$ area ratios range from 0.106 to 0.108 for Co–N–C and from 0.126 to 0.132 for Fe–N–C across the density series (Fig. 4, Fig. S15 and Table S8). The reference calibrations give $R^2$ values of 0.991 for Co and 0.981 for Fe, yielding effective spin values of 1.60–1.67 and 1.71–1.81, respectively. Differences between density levels fall within the estimated uncertainties. These results indicate no resolved density-dependent change in the spin-sensitive XES response under the present measurement and analysis conditions; the absolute effective spin values depend on the reference calibration. Our XES results provide an independent, direct probe of the local spin state across the density series, offering a complementary check to Mössbauer-based assignments.[4] These experimental observations are consistent with the weak dependence of the calculated preferred magnetic moments on site spacing and density across the pristine, HO*-adsorbed, and structural control models (**Fig. 1 and Figs. S6–S8**).

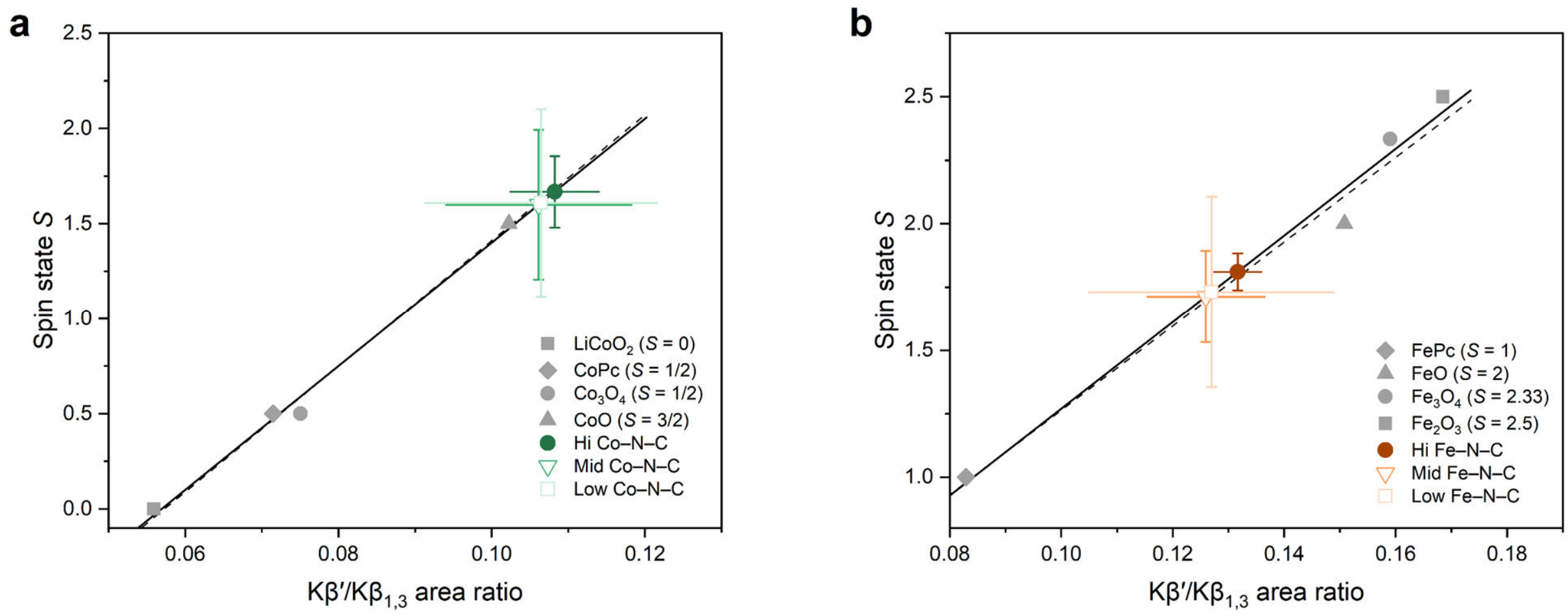

**Fig. 4. Spin-sensitive Kβ XES analysis of density-tuned catalysts.** Effective spin S versus $K\beta'/K\beta_{1,3}$ area ratio for (a) Co and (b) Fe references and the corresponding Hi, Mid and Low catalysts. Grey symbols represent reference compounds with nominal spin values; colored symbols represent the catalysts. Solid and dashed lines denote unweighted and uncertainty-weighted linear calibrations, respectively. Sample error bars represent 1σ uncertainties from Poisson Monte Carlo resampling.

Room-temperature $^{57}Fe$ Mössbauer spectroscopy was further employed to examine the local Fe environments across the density series (Fig. 5). All three spectra were fitted with two quadrupole doublets, D1 and D2, with D1 accounting for 67 ± 4%, 72 ± 3%, and 66 ± 2% of the spectral area in Hi, Mid, and Low Fe–N–C, respectively. Although sample-to-sample variations are present, neither the fitted hyperfine parameters nor the component areas exhibit a monotonic dependence on site density (Table S7). No magnetically split component was resolved at room temperature. These observations complement the XES results and the calculated magnetic-moment trends: the density-dependent ORR selectivity is not accompanied by a corresponding monotonic change in the fitted Mössbauer components. Measurement and fitting details are provided in Supplementary Section 2.6.

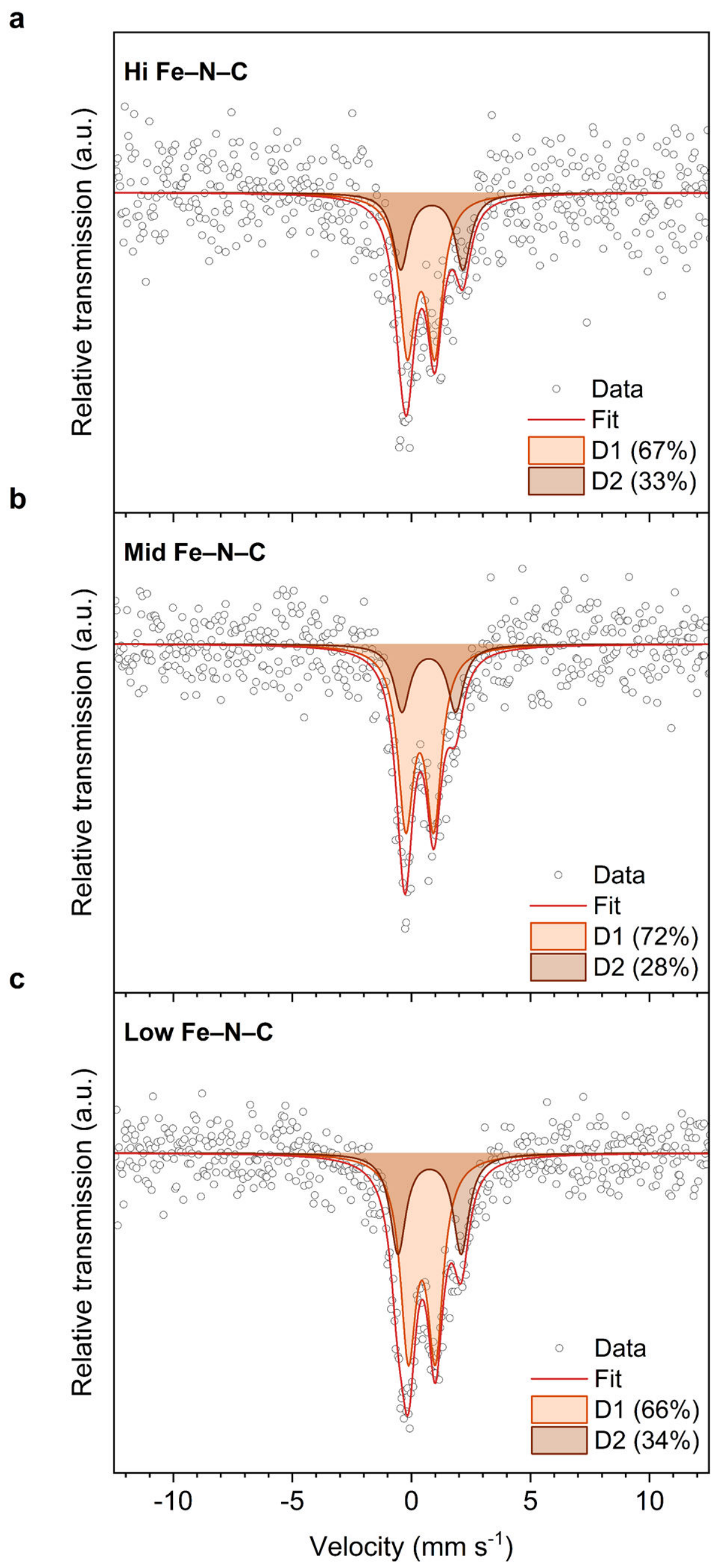

**Fig. 5.** Room-temperature $^{57}Fe$ Mössbauer spectra of Hi, Mid and Low Fe–N–C (transmission geometry, $^{57}Co(Rh)$ source; isomer shifts relative to α-Fe at room temperature). Open circles represent the experimental spectra, red lines the total fits, and the shaded components the fitted doublets D1 and D2. The spectra were fitted with two Lorentzian doublets with a common line width fixed at 0.80 mm $s^{-1}$. D1 ($\delta$ = 0.35–0.44 mm $s^{-1}$, $\Delta E_Q$ = 1.13–1.18 mm $s^{-1}$; 67 ± 4, 72 ± 3 and 66 ± 2 % for Hi, Mid and Low) is assigned to Fe(III)–$N_4$ / low-spin Fe(II)–$N_4$ sites and D2 ($\delta$ = 0.73–0.86 mm $s^{-1}$, $\Delta E_Q$ = 2.26–2.65 mm $s^{-1}$) to high-spin Fe(II)–$N_4$ sites. No magnetically split component is detected.

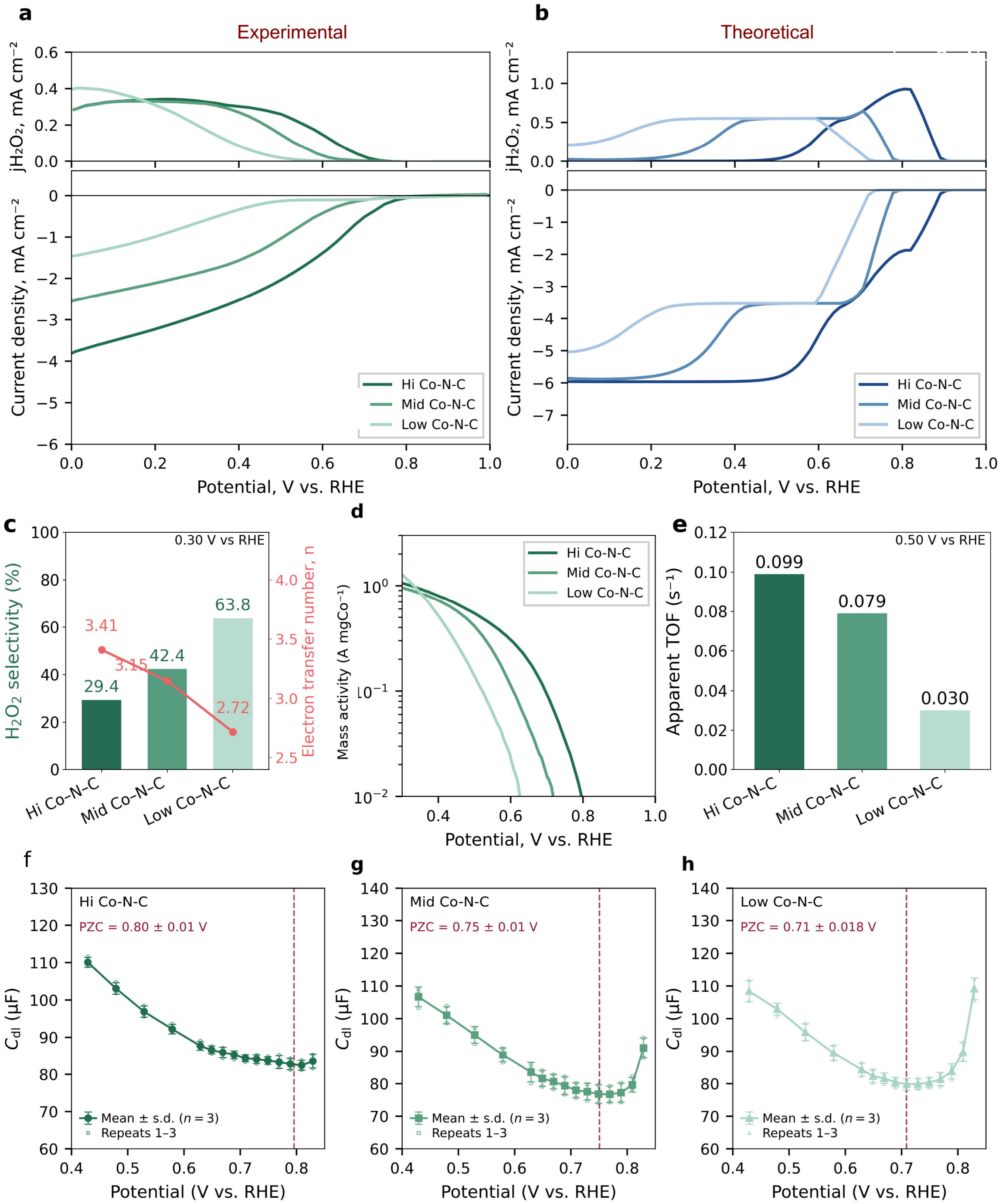

a
Experimental
jH₂O₂, mA cm⁻²
Current density, mA cm⁻²
Potential, V vs. RHE
Hi Co-N-C
Mid Co-N-C
Low Co-N-C
b
Theoretical
jH₂O₂, mA cm⁻²
Current density, mA cm⁻²
Potential, V vs. RHE
Hi Co-N-C
Mid Co-N-C
Low Co-N-C
c
0.30 V vs RHE
$H_2O_2$ selectivity (%)
Electron transfer number, n
3.41
3.15
2.72
29.4
42.4
63.8
Hi Co-N-C
Mid Co-N-C
Low Co-N-C
d
Mass activity (A mgCo⁻¹)
Potential, V vs. RHE
Hi Co-N-C
Mid Co-N-C
Low Co-N-C
e
0.50 V vs RHE
Apparent TOF (s⁻¹)
0.099
0.079
0.030
Hi Co-N-C
Mid Co-N-C
Low Co-N-C
f
Hi Co-N-C
PZC = 0.80 ± 0.01 V
$C_{dl}$ (μF)
Potential (V vs. RHE)
Mean ± s.d. (n = 3)
Repeats 1–3
g
Mid Co-N-C
PZC = 0.75 ± 0.01 V
$C_{dl}$ (μF)
Potential (V vs. RHE)
Mean ± s.d. (n = 3)
Repeats 1–3
h
Low Co-N-C
PZC = 0.71 ± 0.018 V
$C_{dl}$ (μF)
Potential (V vs. RHE)
Mean ± s.d. (n = 3)
Repeats 1–3

**Fig. 6**. **Electrochemical ORR performance of Co–N–C catalysts in 0.1 M HClO4.** (a) RRDE measurements showing ring current density (top, $H_2O_2$ oxidation) and disk current density (bottom) for Hi, Mid, and Low Co–N–C catalysts. (b) Theoretically predicted LSV curves considering PZC in pH-electric-field-coupled microkinetic model. (c) $H_2O_2$ selectivity and electron transfer number (n) at 0.3 V vs. RHE in acidic electrolyte, showing increasing 2-electron selectivity with decreasing site density. (d) Mass activities normalized to Co content, with (e) showing TOF values at 0.5 V vs. RHE: Hi Co–N–C exhibits the highest TOF (0.099 $s^{-1}$), followed by Mid (0.079 $s^{-1}$) and Low (0.03 $s^{-1}$). Double-layer capacitance (Cdl, evaluated at 1 Hz) as a function of potential for (f) Hi, (g) Mid and (h) Low Co–N–C catalysts (mean ± s.d. of three repeated potential series and the individual repeats); the PZC is 0.80 ± 0.01, 0.75 ± 0.01 and 0.71 ± 0.018 V vs RHE, respectively (Table S9).

The ORR performance of the Co–N–C catalysts was then assessed under identical conditions. Their linear sweep voltammetry (LSV) polarization curves collected in an $O_2$-saturated 0.1 M $HClO_4$ electrolyte are shown in Fig. 6a. A clear site density-dependent performance is observed. The catalyst with more active sites shows better ORR activity, with a higher onset potential and a greater limiting current density. The simulated curves capture the qualitative ordering of ORR activity across the density series; their absolute current densities are comparative model outputs (Fig. 6b). As site density decreases, the overall ORR current decreases, whereas $H_2O_2$ selectivity increases. The 4e/2e-ORR pathway preference of these Co–N–C catalysts is further compared. As shown in Fig. 6c, the calculated $H_2O_2$ selectivity ($XH_2O_2$) increases monotonically as site density decreases, i.e., 29.4%, 42.4%, and 63.8% for Hi, Mid, and Low Co–N–C, respectively, at 0.3 V vs. RHE. This trend is accompanied by their reduced electron transfer number (n), which decreases from 3.41 to 3.15 and 2.72. Metal-normalized activities and apparent turnover frequencies (TOFs) were compared using the total metal content determined by ICP-OES (Fig. 6d,e). These quantities assume that all metal atoms contribute equivalently to the site count and do not independently determine the turnover rate per accessible active site. At 0.50 V vs RHE, the apparent TOF values are 0.099, 0.079, and 0.030 $s^{-1}$ for Hi, Mid, and Low Co–N–C, respectively. We further determined the PZC of these catalysts from the potential–double-layer-capacitance (E–$C_{dl}$) correlation obtained from electrochemical impedance measurements, now repeated three times for every catalyst (Fig. 6f–h, Table S9). A characteristic minimum is found for each catalyst, corresponding to its PZC. The PZC decreases from 0.80 ± 0.01 V vs RHE for Hi Co–N–C to 0.75 ± 0.01 V for Mid and 0.71 ± 0.018 V for Low Co–N–C (mean ± s.d., n = 3); the differences between adjacent density levels (0.04–0.05 V) exceed the measurement uncertainty. This trend agrees with the direction predicted by the AIMD-derived values (Fig. 2b), i.e. a lower PZC at lower site density, and confirms that site density modulates the electrochemical boundary conditions at the catalyst–electrolyte interface. A quantitative discrepancy in the absolute PZC values is observed between theory and experiment. This difference is expected, as the calculated PZC corresponds to an idealized M–$N_4$ structural motif embedded in a well-defined carbon lattice, whereas practical M–N–C materials inevitably possess structural heterogeneity, including locally distorted carbon frameworks, various defect types, edge sites, non-uniform coordination environments and surface oxygen functionalities, all of which contribute to the capacitance minimum of the porous electrode.

The ORR performance of the site density varied Fe–N–C catalysts (**Figure S16**) were also studied to assess the generality of the PZC-mediated electrocatalytic activity. The Fe–N–C catalysts exhibit qualitatively similar trends, *i.e.*, higher site density correlates with reduced $H_2O_2$ selectivity, consistent with PZC-mediated mechanism of ORR pathways. The $H_2O_2$ selectivity at 0.3 $V_{RHE}$ follows the expected trend: 2.7%, 13.9%, and 27.6% for Hi, Mid, Low Fe–N–C samples, with corresponding electron transfer numbers of 3.95, 3.72, and 3.45, respectively (**Figure S16c**). Notably, Mid Fe–N–C exhibits the highest apparent TOF of 6.20 $s^{-1}$ at 0.5 $V_{RHE}$, followed by Hi Fe–N–C (3.58 $s^{-1}$) and Low Fe–N–C (0.37 $s^{-1}$), indicating that activity and peroxide selectivity need not exhibit the same dependence on site density. The substantially lower $H_2O_2$ selectivity of Fe–N–C compared to Co–N–C (2.7% vs. 29.4% for the Hi-density samples) reflects the intrinsically stronger *O binding on Fe–$N_4$ sites, which favors the dissociative 4-electron pathway. The PZC of the Fe–N–C catalysts (Fig. S16f–h, Table S9) shows the same trend, 0.79 ± 0.015, 0.75 ± 0.005 and 0.68 ± 0.015 V vs RHE for Hi, Mid and Low Fe–N–C. Complementary measurements were performed to assess electrode-level contributions to the measured peroxide yield. All six catalysts exhibited rapid water uptake without a resolved density-dependent difference at the available temporal resolution (Fig. S18), while $N_2$ physisorption showed broadly comparable pore structures within each series (Fig. S19 and Table S10). Although higher-density electrodes exhibited larger geometric $H_2O_2$-reduction currents, normalization by metal content revealed no systematic enhancement at higher site density (Fig. S20). Thus, the higher peroxide selectivity at lower density is not accompanied by weaker peroxide reduction per unit metal. These externally supplied $H_2O_2$ measurements do not directly quantify peroxide consumption during ORR.

Within the investigated Co–N–C and Fe–N–C series, site density influences the electrochemical boundary conditions at the catalyst–electrolyte interface and the competition between ORR pathways. In this context, the site-density-dependent PZC shift provides a unified and physically grounded explanation for the observed trends in activity and $H_2O_2$ selectivity for both Co–N–C and Fe–N–C catalysts. Compared with prior interpretations that attribute density-dependent performance variations primarily to changes in magnetic moments[4], the PZC-mediated theoretical framework offers a more consistent account of the pronounced activity/selectivity differences observed experimentally.

**Conclusion**

In summary, we do not suggest that spin effects are irrelevant in electrocatalysis. Rather, we suggest that in field-free settings, spin descriptors should be invoked only after validating the magnetic ground state and separating spin-related variables from electrochemical boundary conditions. We show that site-density-dependent ORR activity and selectivity in M–N–C single-atom catalysts are most consistently rationalized by shifts in the PZC, which offer a unified account of the observed trends, whereas the ground-state spin appears largely unchanged across the density range: the calculations and spin-sensitive measurements show comparatively weak variation across the investigated density series. Constrained-magnetization calculations indicate that the ground-state magnetic moments are essentially insensitive to inter-site distance over the range relevant to these materials, whereas explicit-solvent simulations reveal a monotonic decrease in PZC as the metal-site density decreases.

Building on this insight, a PZC-/interfacial-field-aware microkinetic model predicts that lowering the PZC suppresses the $4e^-$ ORR route while promoting the $2e^-$ pathway under acidic conditions. These predictions are corroborated experimentally: both Co–N–C and Fe–N–C catalysts display systematically increasing $H_2O_2$ selectivity with decreasing site density, and the experimentally measured PZC decreases from the high- to the low-density catalysts, in line with the predicted direction. Control experiments show broadly comparable wetting and pore structures, while the higher peroxide selectivity at lower density is not accompanied by weaker $H_2O_2$ reduction per unit metal.

More broadly, our results highlight the importance of rigorous ground-state validation before invoking spin-based descriptors in magnetic-field-free catalysis. The combined constrained-magnetization and explicit-solvent protocol provides a practical framework to disentangle genuine spin effects from correlated electrochemical factors, enabling more reliable and rational design of M–N–C catalysts.

**Data availability**

Computational input files and structural configurations are available at the GitHub repository https://github.com/tohokudizhang/M-N-C_density. The repository includes Quantum ESPRESSO and VASP input files for constrained-magnetization calculations, covering Co–N–C models with varying metal-site densities, Fe–pyrrolic–N models, and $Co_2$–N–C and $Fe_2$–N–C models. It also provides structures sampled from AIMD trajectories for PZC calculations, adsorption-energy configurations, and a Jupyter notebook for analyzing average metal-site distances. The VASP calculation inputs comprise INCAR, POSCAR, and KPOINTS files.

**Acknowledgements**

The authors acknowledge the support provided by JSPS KAKENHI (Nos. JP25H01508, JP25K01737, JP25K17991, and JP24K23069) and National Natural Science Foundation of China (No. 22309109). This work was also supported by the Australian Research Council under the Future Fellowship (FT210100218) and Linkage Project (LP230200886). The authors are grateful for beamtime at the Australian Synchrotron, part of ANSTO, for XAS measurement under project ID 23466 and 27156. The authors acknowledge the XRF facility at Mark Wainwright Analytical Centre (MWAC), UNSW, for XES measurements. The authors acknowledge the Center for Computational Materials Science, Institute for Materials Research, Tohoku University for the use of MASAMUNE-IMR (202512-SCKXX-0218, 202512-SCKXX-0215) and ISSP supercomputers in Tokyo University.

**Author contributions**

D.Z. conceived the theoretical framework and performed the DFT, AIMD, and microkinetic calculations, together with data analysis and interpretation. Z.Y., F.L., Y.L., J.C., X.G., and Y.C. carried out catalyst synthesis and electrochemical measurements and analyzed the experimental data. L.W. supervised the experimental work and contributed to data interpretation. H.L. conceived and supervised the overall project. D.Z., L.W., and H.L. wrote the manuscript with input from all authors.

**Competing interests**

The authors declare no competing interests.